\documentclass{article}
\usepackage{dcase2020,amsmath,graphicx,url,times,booktabs,tabularx}
\usepackage{tikz}
\usepackage{xcolor}
\usetikzlibrary{positioning,intersections,shapes.geometric,fit,arrows,decorations.shapes,decorations.markings}
\usetikzlibrary{backgrounds}

\title{SOUND EVENT LOCALIZATION AND DETECTION BASED ON CRNN USING RECTANGULAR FILTERS AND CHANNEL ROTATION DATA AUGMENTATION}

\name{Francesca Ronchini$^{1}$,
       Daniel Arteaga$^{1,2}$,
       Andrés Pérez-López$^{1,3}$
       }
\address{$^1$ Universitat Pompeu Fabra, Barcelona \\ \{francesca.ronchini01\}@estudiant.upf.edu, \{andres.perez, daniel.arteaga\}@upf.edu\\    
 $^2$ Dolby Iberia, SL, Barcelona \\ 
 $^3$ Eurecat, Centre Tecnologic de Catalunya, Barcelona \\
 }

\begin{document}

\ninept
\maketitle

\begin{sloppy}

\begin{abstract}
Sound Event Localization and Detection refers to the problem of identifying the presence of independent or temporally-overlapped sound sources, correctly identifying to which sound class it belongs, and estimating their spatial directions while they are active. 
In the last years, neural networks have become the prevailing method for Sound Event Localization and Detection task, with convolutional recurrent neural networks being among the most used systems.
This paper presents a system submitted to the  Detection and Classification of Acoustic Scenes and Events 2020 Challenge Task 3. The algorithm consists of a convolutional recurrent neural network using rectangular filters, specialized in recognizing significant spectral features related to the task. In order to further improve the score and to generalize the system performance to unseen data, the training dataset size has been increased using data augmentation. The technique used for that is based on channel rotations and reflection on the xy plane in the First Order Ambisonic domain, which allows improving Direction of Arrival labels keeping the physical relationships between channels. Evaluation results on the development dataset show that the proposed system outperforms the baseline results, considerably improving Error Rate and F-score for location-aware detection.
\end{abstract}

\begin{keywords}
Sound Event Detection, Direction of Arrival estimation, CRNN, First Order Ambisonic, data augmentation, SELD
\end{keywords}

\section{Introduction}
\label{sec:intro}

Sound Event Localization and Detection \textbf{(SELD)} refers to the combined task of Sound Event Detection \textbf{(SED)} and Sound Event Localization \textbf{(SEL)}, whose aim is the recognition of sound sources and their spatial location. In particular, SED requires to identify, instantaneously, the onset and offset of sound events and their correct classification, labeling the event according to the sound class that they belong to. SEL is defined as the estimation of the sound event direction in space with respect to a microphone when an event is active, referred to as Direction-of-Arrival (DOA) estimation. 
Formerly, SED and SEL have been explored as two standalone tasks. Only in recent times they started to be considered jointly. 
In fact, until 2018, the proposed systems considering SELD as a single task were scarce, with only one method based on deep neural network \cite{hirvonen2015classification}. 
In 2018, Advanne et al.~introduced SELDnet \cite{adavanne2018sound}, a convolutional recurrent neural network (CRNN) which simultaneously recognizes, localizes and tracks sound event sources, being the first method to address the localization and recognition of more than two concurrent overlapping sound events. 
The system was proposed as baseline for the Detection and Classification of Acoustic Scenes and Events (DCASE) Challenge 2019 Task 3 \cite{dcase2019task} and for the same task of DCASE Challenge 2020 \cite{dcase2020task}. This year's baseline system includes some modifications inspired by the highest ranked architectures of last year's challenge submissions, among which we can highlight \cite{kapka2019sound}, \cite{cao2019polyphonic} and \cite{zhang2019data}. 
The network receives as input features log-mel spectral coefficients, together with generalized cross-correlation (GCC) for the microphone array (MIC) format, and the acoustic intensity vector in the First Order Ambisonic (FOA) domain. Regarding the network architecture, the model is initially trained with a SED loss only, and then continued with a joint SELD loss. The localization part of the joint loss is masked with the ground truth activations of each class, so that if an event is not active, it does not contribute to the training of the network. 

The methodology proposed in this paper is based on the SELDNet proposed by Adavanne et ~al.~\cite{adavanne2018sound}, including some of the adopted additions in the baseline algorithm of the DCASE 2020 Task 3, such as the use of log-mel spectral coefficients and acoustic intensity vector for the FOA format. However, the system proposed in this paper differs from the baseline system presented for the DCASE 2020 Challenge in several points such as (i) data augmentation, (ii) network architecture and (iii) training loss functions. With respect to (i), -90$^{\circ}$, +90$^{\circ}$ and +180$^{\circ}$ channel rotations of the azimuth angle $\phi$ and its position reflection on the xy and xz planes are used as data augmentation technique, implementing the \textit{16 patterns} spatial augmentation proposed by Mazzon et al. ~\cite{mazzon2019sound}. 
Regarding (ii), the network has been increased, adding 2 convolutional layers. Furthermore, the receptive field has been expanded using rectangular filters (instead of squared ones) in order to make the network able to recognize spectral features relevant for the task. With regard to (iii), we used the same loss functions proposed on the baseline system of last year challenge \cite{adavanne2018sound}. Instead of masked mean square error (used in the baseline system of this year), we used binary cross-entropy loss for SED prediction task and mean square error (MSE) loss for DOA estimation.

Results on the development dataset are evaluated considering the metrics proposed by Mesaros et.~al.~\cite{Mesaros_2019_WASPAA}, which take into account the joint nature of localization and detection. These are the metrics used as evaluation criteria for the challenge.

The paper is structured as follows: Section \ref{sec:method} presents the methodology and the architecture of the proposed system. Section \ref{sec:experiments} describes the experiment setup. Section \ref{sec:res} reports the development results compared with the baseline method. Conclusions and future work are presented in Section \ref{sec:conc}. 

In order to promote reproducibility, the code is available under an open-source license on GitHub\footnote{https://github.com/RonFrancesca/dcase2020-fp}.
\section{METHOD}
\label{sec:method}
The method proposed for the DCASE Challenge Task 3 is based on a enhancement of the system proposed by Advanne et al. \cite{adavanne2018sound}. 

\subsection{Feature Extraction}
This method uses Ambisonic format data. Log mel magnitude spectrogram together with acoustic intensity vector are used as input features for the network. Both are represented in the log mel space to better concentrate the input information of the network, as proposed by Cao et. al in \cite{cao2019polyphonic} and also implemented in the baseline system of the challenge. 

FOA uses four channels to encode spatial information of a sound field, typically denoted as W, X, Y and Z. The channel W corresponds to an omnidirectional microphone recording the sound pressure. The channels X, Y and Z correspond to directional \textit{figure-of-eight} microphones oriented along the components of the homonym cartesian axes, and measure the acoustic velocity of each directional component. The acoustic intensity vector expresses the energy carried by sound waves per area unit in a direction perpendicular to that area, providing DOA information.
The intensity vector is computed as in \cite{cao2019polyphonic}. 

\subsection{The network}
\label{sec:net}
Figure \ref{fig:network} shows the overall architecture of our system with the relative parameter values used in our implementation. 

\tikzstyle{rect} = [rectangle, draw, text centered, minimum width = 9.7em, minimum height = 2em, align=center]
\tikzstyle{rect_large} = [rectangle, draw, text centered, minimum width = 20em, minimum height = 2em, align=center]

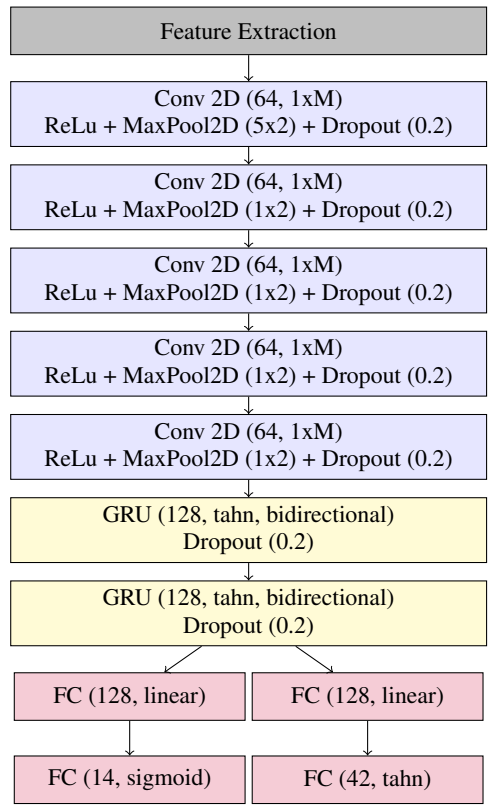
\begin{figure}[h]
\begin{center}
\begin{tikzpicture}[node distance = 3.5em, auto, align=center]
\centering

    \node (osc)     [rect_large, fill=lightgray] {Feature Extraction};
    \node (conv1)   [rect_large, fill=blue!10, below of = osc] {Conv 2D (64, 1xM) \\
    ReLu + MaxPool2D (5x2) + Dropout (0.2)};
    \node (conv2)   [rect_large,  fill=blue!10, below of = conv1] {Conv 2D (64, 1xM) \\
    ReLu + MaxPool2D (1x2) + Dropout (0.2)};
    \node (conv3)   [rect_large,  fill=blue!10, below of = conv2] {Conv 2D (64, 1xM) \\
    ReLu + MaxPool2D (1x2) + Dropout (0.2)};
    \node (conv4)   [rect_large,  fill=blue!10, below of = conv3] {Conv 2D (64, 1xM) \\
    ReLu + MaxPool2D (1x2) + Dropout (0.2)};
    \node (conv5)   [rect_large,  fill=blue!10, below of = conv4] {Conv 2D (64, 1xM) \\
    ReLu + MaxPool2D (1x2) + Dropout (0.2)};
    \node (GRU1)    [rect_large, fill=yellow!20, below of = conv5] {GRU (128, tahn, bidirectional) \\ Dropout (0.2)};
    \node (GRU2)    [rect_large, fill=yellow!20, below of = GRU1] {GRU (128, tahn, bidirectional) \\ Dropout (0.2)};
    \node (FC1)     [rect, xshift=-5em, fill = red!80!blue!20, below of = GRU2] {FC (128, linear)};
    \node (FC2)     [rect, xshift= 5em, fill=red!80!blue!20, below of = GRU2] {FC (128, linear)};
    \node (fc-sed)  [rect, fill = red!80!blue!20, below of = FC1] {FC (14, sigmoid)};
    \node (fc-doa)  [rect, fill=red!80!blue!20, below of = FC2] {FC (42, tahn)};

    \draw [->] (osc)  ->  (conv1);
    \draw [->] (conv1)  ->  (conv2);
    \draw [->] (conv2)  ->  (conv3);
    \draw [->] (conv3)  ->  (conv4) ;
    \draw [->] (conv4)  ->  (conv5) ;
    \draw [->] (conv5)  ->  (GRU1) ;
    \draw [->] (GRU1)  ->  (GRU2) ;
    \draw [->] (GRU2)  ->  ([xshift=-2em] FC1);
    \draw [->] (GRU2)  ->  ([xshift= 2em] FC2);
    \draw [->] (FC1)  ->  (fc-sed);
    \draw [->] (FC2)  ->  (fc-doa);

\end{tikzpicture}
\end{center}

\caption{The proposed network architecture.}
\label{fig:network}
\end{figure}

The proposed system is based on the SELDNet introduced by  Adavanne et al.~in \cite{adavanne2018sound}, with some modifications. In a similar manner, it features a CRNN network using Gated Recurrent Units (GRU) as recurrent layers. This is followed by two parallel branches of Fully Connected (FC) layers, one for SED and one for DOA estimation, sharing weights along time dimension. The first FC layer of both branches uses linear activation, while the last FC layer of each branch uses a different activation function according to the task. The last FC layer in the SED branch contains 14 nodes using sigmoid activation (one node for each sound event class to be detected), while the last FC layer in the DOA branch consists of 42 nodes using $\tanh$ activation (each of the sound event classes is represented by 3 nodes relative to the 3-dimensional sound event location). We use binary cross-entropy as loss function for the SED branch and mean square error (MSE) loss for DOA estimation branch, keeping the two branches separated. 

Regarding the differences respect to the baseline in this implementation, firstly, we added 2 CNN blocks in order to help the network learn more features, increasing the number of CNN blocks from 3 to 5. Each CRNN block consists of a convolutional layer with rectified linear unit (ReLU) activation, batch normalization to normalize the activation output, and MaxPooling along frequency axis to reduce the dimensionality. Although adding layers to a neural network usually helps it to learn more features, it has the disadvantage of leading to possible overfitting, especially when the training dataset size is small as in this case. To prevent overfitting, we use Dropout in each convolutional block, after reducing the dimensionality. 

Secondly, we used rectangular filters instead of squared ones, mainly inspired by Pons et al.~\cite{pons2016experimenting}. In the mentioned paper, the authors studied how filter shapes can help to proper model CNN motivated by musical aspects, achieving positive results in music classifications. Filters used in convolutional layers are principally inspired by image processing literature, typically being small and squared (3x3 or 5x5). Considering that, in the audio domain, filter dimensions corresponds to time and frequency axes, wider filters may be capable of learning longer temporal dependencies in the audio, while higher filters may be capable of learning more frequency features. 
We propose the same concept, applying it to sound event detection. We used rectangular filters of shape ${1xM}$, being 1 the time dimension and the M the frequency dimension. We hypothesize that setting the time dimension to 1 and increasing the frequency dimension, taking into consideration nearly all the mel-bands, would lead the network to better model the frequency dimensions, helping the system to learn the presence or absence of an event, and consequently improving the metrics for location-aware detection. 
Such horizontal filters increase the receptive field only on frequency dimension. Anyway, the temporal information is taken care by the recurrent GRU layers. 

The last addition is the use of data augmentation as described in Section 2.3, expanding the number of DOA represented in it and consequently raising the scores related to classification-dependent localization metrics. 

\subsection{Data augmentation}
\label{sec:data}
With the aim of improving the results and to reduce system overfitting, the training dataset size has been increased using data augmentation based on channel rotations and reflection on the \textit{xy} plane in the FOA domain. 
Among others, this system implements the \textit{16 patterns} technique proposed for the first time by Mazzon et al.~in \cite{mazzon2019sound}, with some small changes. This approach allows to increase DOA combination and correctly compute the corresponding ground truth DOA
labels of the augmented data. Moreover, a relevant advantage of this method is the possibility to be applied regardless of the number of overlapping sound sources \cite{mazzon2019first}, which makes it an easy and straightforward data augmentation method. 
We augmented the data following the transformations suggested in \cite{mazzon2019first}, considering only \textit{channel swapping} and \textit{channel sign inversion}. The suggested data manipulations correspond to rotations of 0, -90$^{\circ}$, +90$^{\circ}$, and +180$^{\circ}$ of the azimuth angle $\phi$ and its reflection with respect to the xz plane, leading to 8 rotations around the z axis, and a reflection with respect to the xy plane (considering the opposite elevation angle), for a total of 15 new patterns plus the original one. 
Figure \ref{fig:intensity} shows an example of channel rotation on intensity vector, after applying a reflection with respect to xy plane. The figure shows the reflection of channel Z. The reader is referred to \cite{mazzon2019first} for further details. 
In \cite{mazzon2019sound}, Mazzon et al.~compute the augmented dataset in time domain and extract the features offline for each of the augmented signals. All the possible transformations are computed offline, and the data generator randomly chooses one of them at each iteration. 
In our system we implemented only 15 patterns, without considering the original one as data augmentation pattern. We also implemented the data augmentation offline, but, for memory reasons, instead of computing all the transformations for each audio file, we randomly select one out of the 15 patterns to augment the data during the feature extraction process. The pattern selected for a particular audio file is also used for augmenting the corresponding label. All the new generated files, together with the original ones, are used to train the network. 

\begin{figure}[t]
\centering
\includegraphics[width=0.5\textwidth]{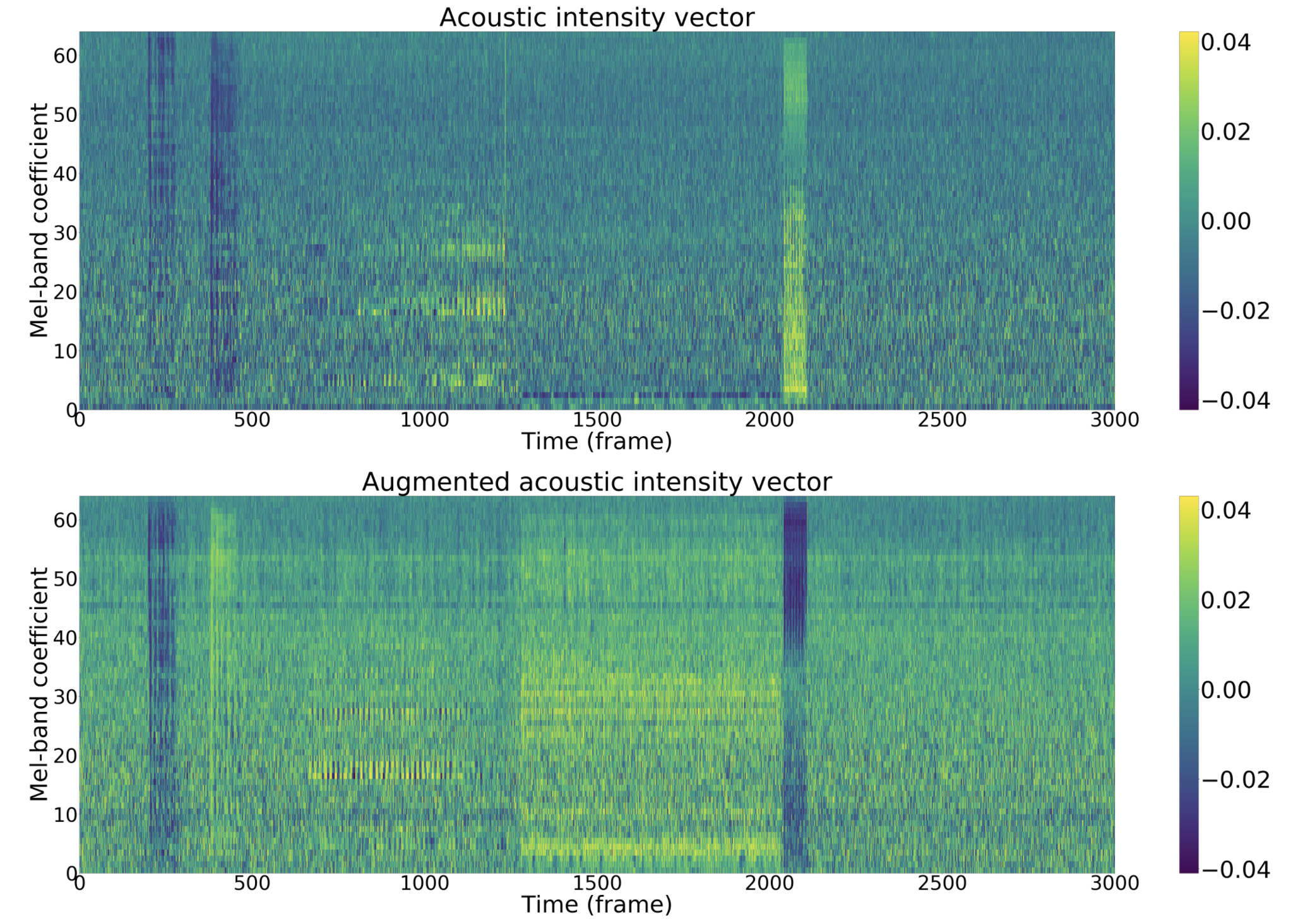}
\caption{Example of augmented intensity vector. Pattern applied: reflection with respect to xy plane. The image shows the Z-component of the intensity vector, before and after applying a reflection on the Z-channel. The horizontal axis represents time (expressed in frames) and the vertical axis mel-band coefficients.}
\label{fig:intensity}
\end{figure}

\subsection{Hyper-parameters}
All the dataset audio files are sampled at 24kHz. For the STFT, we used a 960 point Hanning window with a 50\% hop size. Considering that the temporal resolution of the label is 100 ms, we interpolated the sub-frames of 20ms as suggested in the baseline system. The number of mel-band filter is set to 64. 

The training development set has been increased from 400 to 1200 files using the aforementioned augmentation technique.  
With regard to the optimization technique, we used Adam method \cite{kingma2014adam}. 

A sound event is considered to be active, and its respective DOA estimation considered, if the SED output exceeds a threshold of 0.5. 

\section{Experiments}
\label{sec:experiments}
The dataset used for the evaluation of the system is the one provided for the DCASE 2020 Challenge Task 3: TAU-NIGENS Spatial Sound Events 2020 \cite{politis2020dataset}. 
The dataset consists of sound samples belonging to 14 different sound classes, which have been convolved with real spatial room impulse responses (RIRs). The dataset is presented in two spatial recording formats, MIC and FOA, and it is divided between development and evaluation set. The results presented in this paper are based on the FOA development set of the dataset, which consists of 6 predefined splits. We followed the same specification given in the task description, using split 1 for testing, split 2 for evaluation and split 3-6 for training. Further information regarding the dataset can be found at \cite{politis2020dataset}. 

The network predictions have been evaluated considering the joint nature of localization and detection, as proposed in \cite{Mesaros_2019_WASPAA}. In particular, $ER\textsubscript{20$^{\circ}$}$ and $F\textsubscript{20$^{\circ}$}$ are related to the SED task and they are location-dependent. A prediction is considered true positive only if it is under a distance threshold of {20$^{\circ}$} from the reference. $LE\textsubscript{CD}$ (localization error) and $LR\textsubscript{CD}$ (localization recall) are related to DOA estimation, being classification-dependent, which means they are going to be calculated only between sounds with the same label in each frame. 
All metrics are computed in one-second non-overlapping frames. For more information about the evaluation metrics refer to \cite{Mesaros_2019_WASPAA}. 

Several architecture configurations, filter dimensions and data augmentation techniques have been explored before reaching the network architecture described in Section \ref{sec:net}. It is possible to consider the full experiment as the combination of two minor sub-experiments. The first one has concentrated on finding the rectangular filter shape which gives the best performance. The second has been focused on considering different data augmentation methods to select the most appropriate technique. 

During the first experiment, we explored different rectangular shapes in order to understand the filters size which gives the best performance and helps the network to properly learn frequency features and correctly detect events. To do so, the results of the baseline system, which uses rectangular filters of shape 3x3, have been compared with the results of the proposed system using rectangular filters of shape $TxM$ of size 1x46, 1x48, 2x48, 3x48, 1x50, 1x52, 1x54, 1x56, and completely dense filters of dimension 1x64 (being 64 the size of mel-band filter). We supposed that the use of rectangular filters would lead the network to better model the frequency features. 
We also tested filters of shape $2x48$ and $3x48$ to investigate if increasing the time dimension would facilitate the network to learn temporal information. 

After selecting the filter which best performs, different data augmentation techniques have been considered to increase the SELD score and to reduce overfitting. In this second step, the research has concentrated on three data augmentation techniques: time stretching, pitch shifting and channel rotations. Each technique has been independently explored.
Regarding time stretching, we considered a stretch factor  between 0.9 and 1.1. Regarding pitch shifting, we considered the sound in a range between -2 and 2 semitones, considering only integer values. For each augmentation technique, two values within the chosen ranges have been randomly selected for each audio file, increasing the dataset from 400 to 1200 files. More details regarding channel rotation are given in Section \ref{sec:data}.

Across all the experiments, the batch size has been set to 128, and the systems have been trained for 50 epochs at most. An early stopping strategy has been implemented, stopping the training if the validation loss does not improve for 20 epochs.

\section{Results and discussion}
\label{sec:res}

\begin{table}[!t]
 \begin{tabularx}{\columnwidth}{lccccc}
  \toprule
  Filter shape & ER\textsubscript{20$^{\circ}$} & F\textsubscript{20$^{\circ}$} & LE\textsubscript{CD} & LR\textsubscript{CD} & SELD \\
  \midrule
  Baseline (3x3) & 0.72  & 37.4\%    & 22.8$^{\circ}$    & 60.7\%    & 0.47  \\
  1x46          & 0.69  & 40.1\%    & 21.7$^{\circ}$    & 62.3\%    & 0.45   \\
  \textbf{1x48} & \textbf{0.69}  & \textbf{40.3\%}  & \textbf{20.9$^{\circ}$}    & \textbf{62.4\%}    & \textbf{0.45} \\
  1x50         & 0.70  & 40.4\%    & 21.1$^{\circ}$    & 61.1\%    & 0.45 \\
  1x52          & 0.70  & 39.9\%    & 21.6$^{\circ}$    & 61.4\%    & 0.45 \\
  1x54          & 0.72  & 37.1\%    & 21.7$^{\circ}$    & 59.7\%    & 0.47 \\
  1x56          & 0.72  & 38.2\%    & 21.2$^{\circ}$    & 59.4\%    & 0.47 \\
  1x64          & 0.72  & 38.1\%    & 23.1$^{\circ}$    & 61.4\%    & 0.46 \\ 
  \textbf{2x48}          & \textbf{0.70}  & \textbf{40.6\%}    & \textbf{20.8$^{\circ}$}   & \textbf{61.1\%}    & \textbf{0.45} \\
  3x48          & 0.75  & 36.0\%    & 23.1$^{\circ}$    & 58.6\%    & 0.48 \\
 \bottomrule
 \end{tabularx}
 \caption{Evaluation results on development set for first stage of the experiment using different filters shapes.}
 \label{tab:results_filter}
\end{table} 

Tables \ref{tab:results_filter} and \ref{tab:results_data} report the evaluation results of the first and second stage of the study, respectively. In particular, Table \ref{tab:results_filter} shows the evaluation results for the development dataset on the testing split using different rectangular filters shapes. Table \ref{tab:results_data} details the results on the same dataset, using 1x48 filters with different data augmentation methods. 
In both tables, the results are compared with the baseline system. 

Table \ref{tab:results_filter} reports the results related to the first sub-experiment. 
In order to select the filters shapes which give the best results, several sizes has been tested. As it is possible to observe, between all the filters of size $1xM$, the best results are reached with filter shapes of 1x48 and 1x50. Among those, almost all the tested rectangular shapes outperfom the baseline results. The only filter which does not improve the metrics is the 1x54, which is otherwise comparable to it. Between the two filter shapes which better perform, we selected the 1x48 to continue the experiment. In particular, the selection of the 1x48 filters instead of 1x50 is based on the fact that, while the difference on sound event detection metrics is negligible, localization-aware classification scores for the first shape are better than the second one. 
The table also report the results of filters's size 2x48 and 3x48 (on the same table for space limitations), which have been tested (after selecting 1x48 filter shape) to explore their relation with temporal information. As it possible to observe, increasing the receptive field also in time dimension does not help improve the results. Comparing 2x48 and 1x48, the difference in terms of results is neglectable, while increasing the time dimension also increase the training time of the network and its size. For this reasons, we selected the 1x48 as the best performant and as the final filter size to follow to study with. 
This shows that using rectangular filters increasing the receptive field only in frequency dimension help the network to better model frequency features while GRU layers take care of the temporal information. Moreover, it also highlights the importance of understanding the training datasets in order to properly design the model architecture for a determined task. 

The second sub-experiment is based on the exploration of several data augmentation techniques with the aim to increase the score and prevent overfitting. We focused the study on three methods: time stretching, pitch shifting and channel rotation. Results are shown in Table \ref{tab:results_data}. As it possible to observe, channel rotation is the only method that outperforms the baseline results, substantially increasing location-aware detection scores. This might be explained by the fact that channel rotations maintains the physical relations between channels, making the manipulations of sound more realistic. Moreover, the noticeable improvement of SED-related scores ($ER$ and $F-score$) could be explained by the joint nature of localization and detection of the evaluation metrics. In fact, the results suggest that channel rotation augmentation technique increases the DOA information, helping the network to better localize sound sources. It is possible that, without the expansion of the dataset size, the same sound source would have been considered as active, but the system would have been wrong in localizing it, considering the prediction wrong. By augmenting the data and by giving the network more DOA examples for training, the algorithm's results improve in a substantial way. 

\begin{table}[!t]
 \begin{tabularx}{\columnwidth}{l@{\hskip 0.2in}c@{\hskip 0.2in}c@{\hskip 0.2in}c@{\hskip 0.2in}c@{\hskip 0.2in}c}
  \toprule
  Method & ER\textsubscript{20$^{\circ}$} & F\textsubscript{20$^{\circ}$} & LE\textsubscript{CD} & LR\textsubscript{CD} & SELD  \\
  \midrule
  Baseline          & 0.72  & 37.4\%    & 22.8$^{\circ}$    & 60.7\%    & 0.47 \\
  TS  & 0.86  & 22.8\%    & 27.9$^{\circ}$    & 51.0\%    & 0.57 \\
  PS    & 0.86  & 22.3\%    & 28.7$^{\circ}$    & 51.0\%    & 0.57 \\
  \textbf{CR}  & \textbf{0.59}  & \textbf{50.6\%}    & \textbf{17.6$^{\circ}$}    & \textbf{66.2\%}    & \textbf{0.38} \\
  \bottomrule
 \end{tabularx}
 \caption{Evaluation results on development for the second stage of the experiment using different data augmentation techniques. TS stands for time stretching, PS stands for pitch shifting and CR stands for channel rotations. }
 \label{tab:results_data}
\end{table}

\section{Conclusions and Future Work}
\label{sec:conc}
This paper describes a system submitted for the DCASE Challenge 2020 Task 3. The method is based on SELDNet presented by Adavanne et. al~ \cite{adavanne2018sound}, with some differences. Data augmentation based on Ambisonic rotations, network architecture and training loss functions are the main changes. 
The main contribution of the proposed system is the usage of rectangular filters, highlighting the importance of modeling the network architecture according to the dataset used for training. Data augmentation also helped to increase the evaluation score, especially ER\textsubscript{20$^{\circ}$} and F\textsubscript{20$^{\circ}$} related to the SED task. 
The proposed system considerably outperforms the state-of-the-art method presented as baseline, significantly increasing the location-dependent metrics related to SED task.

Future work will include further investigation of different data augmentation techniques based on preserving physical relations between channels, such as random rotation matrices to rotate the acoustic scene of random angles, and explore rectangular filter shapes performances on different dataset formats.
\newpage

\bibliographystyle{IEEEtran}
\bibliography{refs}

\begin{thebibliography}{10}
\providecommand{\url}[1]{#1}
\def\UrlFont{\rmfamily}
\providecommand{\newblock}{\relax}
\providecommand{\bibinfo}[2]{#2}
\providecommand\BIBentrySTDinterwordspacing{\spaceskip=0pt\relax}
\providecommand\BIBentryALTinterwordstretchfactor{4}
\providecommand\BIBentryALTinterwordspacing{\spaceskip=\fontdimen2\font plus
\BIBentryALTinterwordstretchfactor\fontdimen3\font minus
  \fontdimen4\font\relax}
\providecommand\BIBforeignlanguage[2]{{%
\expandafter\ifx\csname l@#1\endcsname\relax
\typeout{** WARNING: IEEEtran.bst: No hyphenation pattern has been}%
\typeout{** loaded for the language `#1'. Using the pattern for}%
\typeout{** the default language instead.}%
\else
\language=\csname l@#1\endcsname
\fi
#2}}

\bibitem{hirvonen2015classification}
T.~Hirvonen, ``Classification of spatial audio location and content using
  convolutional neural networks,'' in \emph{Audio Engineering Society
  Convention 138}.\hskip 1em plus 0.5em minus 0.4em\relax Audio Engineering
  Society, 2015.

\bibitem{adavanne2018sound}
S.~Adavanne, A.~Politis, J.~Nikunen, and T.~Virtanen, ``Sound event
  localization and detection of overlapping sources using convolutional
  recurrent neural networks,'' \emph{IEEE Journal of Selected Topics in Signal
  Processing}, vol.~13, no.~1, pp. 34--48, 2018.

\bibitem{dcase2019task}
\url{http://dcase.community/challenge2019/task-sound-event-localization-and-detection}.

\bibitem{dcase2020task}
\url{http://dcase.community/challenge2020/task-sound-event-localization-and-detection}.

\bibitem{kapka2019sound}
S.~Kapka and M.~Lewandowski, ``Sound source detection, localization and
  classification using consecutive ensemble of crnn models,'' \emph{arXiv
  preprint arXiv:1908.00766}, 2019.

\bibitem{cao2019polyphonic}
Y.~Cao, Q.~Kong, T.~Iqbal, F.~An, W.~Wang, and M.~D. Plumbley, ``Polyphonic
  sound event detection and localization using a two-stage strategy,''
  \emph{arXiv preprint arXiv:1905.00268}, 2019.

\bibitem{zhang2019data}
J.~Zhang, W.~Ding, and L.~He, ``Data augmentation and prior knowledge-based
  regularization for sound event localization and detection,'' Tech. Report of
  Detection and Classification of Acoustic Scenes and Event, Tech. Rep., 2019.

\bibitem{mazzon2019sound}
L.~Mazzon, M.~Yasuda, Y.~Koizumi, and N.~Harada, ``Sound event localization and
  detection using foa domain spatial augmentation,'' in \emph{Proc. of the 4th
  Workshop on Detection and Classification of Acoustic Scenes and Events
  (DCASE)}, 2019.

\bibitem{Mesaros_2019_WASPAA}
A.~Mesaros, S.~Adavanne, A.~Politis, T.~Heittola, and T.~Virtanen,
  ``\BIBforeignlanguage{English}{Joint measurement of localization and
  detection of sound events},'' in \emph{\BIBforeignlanguage{English}{IEEE
  Workshop on Applications of Signal Processing to Audio and Acoustics
  (WASPAA)}}, New Paltz, NY, Oct 2019, accepted.

\bibitem{pons2016experimenting}
J.~Pons, T.~Lidy, and X.~Serra, ``Experimenting with musically motivated
  convolutional neural networks,'' in \emph{2016 14th international workshop on
  content-based multimedia indexing (CBMI)}.\hskip 1em plus 0.5em minus
  0.4em\relax IEEE, 2016, pp. 1--6.

\bibitem{mazzon2019first}
L.~Mazzon, Y.~Koizumi, M.~Yasuda, and N.~Harada, ``First order ambisonics
  domain spatial augmentation for dnn-based direction of arrival estimation,''
  \emph{arXiv preprint arXiv:1910.04388}, 2019.

\bibitem{kingma2014adam}
D.~P. Kingma and J.~Ba, ``Adam: A method for stochastic optimization,''
  \emph{arXiv preprint arXiv:1412.6980}, 2014.

\bibitem{politis2020dataset}
\BIBentryALTinterwordspacing
A.~Politis, S.~Adavanne, and T.~Virtanen, ``A dataset of reverberant spatial
  sound scenes with moving sources for sound event localization and
  detection,'' \emph{arXiv e-prints: 2006.01919}, 2020. [Online]. Available:
  \url{https://arxiv.org/abs/2006.01919}
\BIBentrySTDinterwordspacing

\end{thebibliography}

%
%
%
%
%
%
%
%
%

\end{sloppy}
\end{document}